\documentclass[]{aastex7}

\usepackage{xcolor}
\usepackage{booktabs}
\usepackage{multirow}
\usepackage{amsmath}
\graphicspath{{./}{figures/}}
\usepackage{graphicx}

\shorttitle{Ly$\alpha$ RT modeling for MUSE LAEs}
\shortauthors{Yu et al.}

\begin{document}

\title{Ly$\alpha$ radiative transfer modeling for 163 MUSE Ly$\alpha$-emitting galaxies at $z=$3--6}

\author[0009-0001-3373-9747]{Sangeun Yu}
\affiliation{Department of Astronomy, Space Science and Geology, Chungnam National University, 99 Daehak-ro, Yuseong-gu, Daejeon 34134, Republic of Korea}
\email[]{dbse1025@gmail.com}

\author[0000-0002-4362-4070]{Hyunmi Song}
\affiliation{Department of Astronomy and Space Science, Chungnam National University, 99 Daehak-ro, Yuseong-gu, Daejeon 34134, Republic of Korea}
\email[show]{hmsong@cnu.ac.kr}

\author[0000-0001-9561-8134]{Kwang-il Seon}
\affiliation{Korea Astronomy $\&$ Space Science Institute, 776, Daedeokdae-ro, Yuseong-gu, Daejeon 34055, Republic of Korea}
\affiliation{Department of Astronomy and Space Science, University of Science and Technology, 217, Gajeong-ro, Yuseong-gu, Daejeon 34113, Republic of Korea}
\email[]{kiseon@kasi.re.kr}

\begin{abstract} \nolinenumbers
We utilized Ly$\alpha$ radiative transfer calculations from \citet{Song2020} to investigate the properties of extended Ly$\alpha$ halos around star-forming galaxies in the \textit{Hubble} Ultra Deep Field, observed by the Multi-Unit Spectroscopic Explorer.
Expanding on the work of \citet{Song2020}, which was limited to eight galaxies, we derived best-fit models for a significantly larger sample of 163 galaxies, which successfully reproduced both their Ly$\alpha$ spectra and surface brightness profiles (SBPs).
These best-fit models suggest a broad medium distribution surrounding each galaxy, with low expanding velocities at large radii.
This conclusion could not have been drawn from modeling either the spectrum or SBP alone, but only through simultaneous modeling of both.
Our correlation analysis between observables and model parameters reveals that the spatial extent of Ly$\alpha$ halos is primarily determined by the extents of the medium and the source, while the spectral peak shift and full width at half maximum are governed mainly by optical depth, with the velocity structure of the medium playing a secondary yet non-negligible role.
The fact that various correlations derived from the full set of models and those from the best-fit subset can differ significantly highlights the complex and interdependent nature of Ly$\alpha$ radiative transfer.
All model parameters interact to shape the observed Ly$\alpha$ features in a non-trivial way.
\end{abstract}

\section{Introduction}

Ly$\alpha$ is one of the strongest nebular emission lines, with its emission mechanism directly linked to star formation activities, making it commonly observed in star-forming galaxies \citep{Barnes2014,Dijkstra2014,Hayes2015}. Therefore, it serves as a key observable for studying galaxies or cosmic epochs with active star formation. However, because Ly$\alpha$ is a resonant line of hydrogen, the most abundant element in the universe and the transition with one of the largest cross sections, it inherently comes with complexities. These complexities arise from the fact that the scattering and absorption of Ly$\alpha$ are intricately influenced by various physical properties of the medium it passes through.
In other words, it is difficult to easily infer the physical properties of the medium or the intrinsic properties from Ly$\alpha$ observables, e.g., the emerging spectral shape and surface brightness distribution.

The complex Ly$\alpha$ radiative transfer has been studied for various medium distributions using Monte Carlo numerical methods. Starting with the simplest cases like uniform, static, semi-infinite slabs \citep[for which analytic solutions can be derived][]{Neufeld1990}, a range of media has been considered, including spherically symmetric halos or shells with various motions \citep[including static ones][]{Zheng_MiraldaEscude2002,Ahn2003,Dijkstra2006,Verhamme2006,Schaere2011,Gronke2015,Yang2016,Song2020,Seon2020,Li_Gronke2022}, clumpy media \citep{Hansen_Oh2006,Gronke_Dijkstra2016,Gronke_etal2017,Chang2023}, and those from hydrodynamical simulations \citep{Laursen_SommerLarsen2007,Kollmeier2010,Zheng2011,Verhamme2012,Behrens_Braun2014,Smith2015,Smith2019,Smith2022,Garel2021,Blaizot2023}. Among these, media from hydrodynamical simulations, which are considered the most realistic though, make it difficult to determine which physical quantities (or model parameters) are involved in Ly$\alpha$ radiative transfer and how influential they are.  This is because it is not easy to control each physical parameter that characterizes the sources and media.  As a result, even though they are less realistic, simplified models such as halos, shells, and somewhat simplified clumpy media remain useful for understanding Ly$\alpha$ radiative transfer.

By considering various combinations of physical parameters that are involved in Ly$\alpha$ radiative transfer and pinning down one that best reproduces observed data, one can place constraints on these parameters and thus have understanding of the Ly$\alpha$ radiative transfer process in observed galaxies. The observables being reproduced are primarily spectra, and previous studies have demonstrated that the models mentioned above have been able to successfully replicate a range of observed spectral shapes \citep[e.g.,][]{Yang2017,Gronke2017}. However, caution is required when interpreting the best-fitting model due to the degeneracy between parameters. The degeneracy is often found between the hydrogen column density (or equivalently the nominal optical depth $\tau_0$) and velocity profile of the medium, both of which contribute to the actual optical depth for individual photons. To break such degeneracies, it may be useful to consider other observables. One of the most straightforward options is the surface brightness distribution \citep{Song2020}, and going further, polarization characteristics could also be considered \citep{Seon2022,Chang2023}. However, since there are currently limited high signal-to-noise polarization data available \citep[e.g.,][]{Hayes2011,Bower2011,Prescott2011,You2017,Kim2020}, the surface brightness distribution is considered more practical for now.

Obtaining both spectra and surface brightness distributions simultaneously is most efficiently achieved through Integral Field Unit (IFU) observations, with high-redshift Lyman-alpha emitters (LAEs) observed by Multi-Unit Spectroscopic Explorer (MUSE) serving as an excellent example \citep{Bacon2010,Bacon2014}.
MUSE, developed as the second instrument for the ESO-VLT, has been in operation since 2014. It possesses the sensitivity and stability to perform observations to match the depth of \textit{Hubble} (Ultra) Deep Fields, along with high spatial and spectral resolution, large multiplexing capabilities, and extensive spectral coverage. This allows the study of galaxy kinematics, metallicity, and environment or clustering through precise redshift measurements that broadband photometric data cannot provide. Particularly, as mentioned earlier, Ly$\alpha$ photons undergo significant scattering within the medium, leading to diffusion in frequency and spatial domains. This necessitates high spectral resolution and photometric sensitivity to effectively observe and analyze their properties. 
MUSE meets these requirements, making it ideal for studying the Ly$\alpha$ characteristics of galaxies.

\citet{Leclercq2017} studied the properties of Ly$\alpha$ halos using MUSE data of 145 star-forming galaxies at redshifts $3 \leq z \leq 6$. This study followed up on the work by \citet{Wisotzki2016}, which focused on 26 galaxies, significantly increasing the sample size and improving the limiting surface brightness sensitivity. In the work by \citet{Song2020}, the spectra and surface brightness profiles of eight galaxies of \citet{Leclercq2017}, which are digitized from the figure in \citet{Leclercq2017}, were modeled using the Monte Carlo radiative transfer calculations for an expanding halo. They demonstrated that the observed spectra and surface brightness profiles could be successfully reproduced with the rather simple model, significantly alleviating parameter degeneracies by simultaneously reproducing both observables (see their Section 4.2).
This study aimed to extend previous study by reproducing the spectra and surface brightness profiles of most galaxies in the \citet{Leclercq2017}. Based on the increased sample size, we revisited the correlations between the observables and model parameters examined in prior studies and explored the properties of high-redshift star-forming galaxies revealed by the modeling results.

This paper is organized as follows.
In Section \ref{sec:obs}, we present the MUSE IFU data of \citet{Leclercq2017}.
Section \ref{sec:mod} describes the Ly$\alpha$ radiative models calculated by \citet{Song2020} and the modeling for our galaxy sample. 
We then present and discuss the results of the modeling in Section \ref{sec:res}.
Finally, we conclude with a summary in Section \ref{sec:sum}.

\section{Observed spectra and surface brightness profiles from MUSE IFU data}\label{sec:obs}

We utilized Ly$\alpha$ spectra and surface brightness profiles of star-forming galaxies at redshifts $3 \leq z \leq 6$ in \citet{Leclercq2017} that were derived from the MUSE IFU data in the \textit{Hubble} Ultra Deep Field. 
Starting from 775 LAEs with reliable redshift measurements, they left 184 LAEs after removing those in pair, close to the survey edges, and/or under detection limits considering the noise and the point spread function (PSF) of the MUSE data. 
Upon visual inspection of the spectrum of these LAEs, we found that 19 objects exhibit spectra with double peaks and the rest spectra with red peak only.

\begin{figure}[t]
    \centering
    \includegraphics[width=15cm]{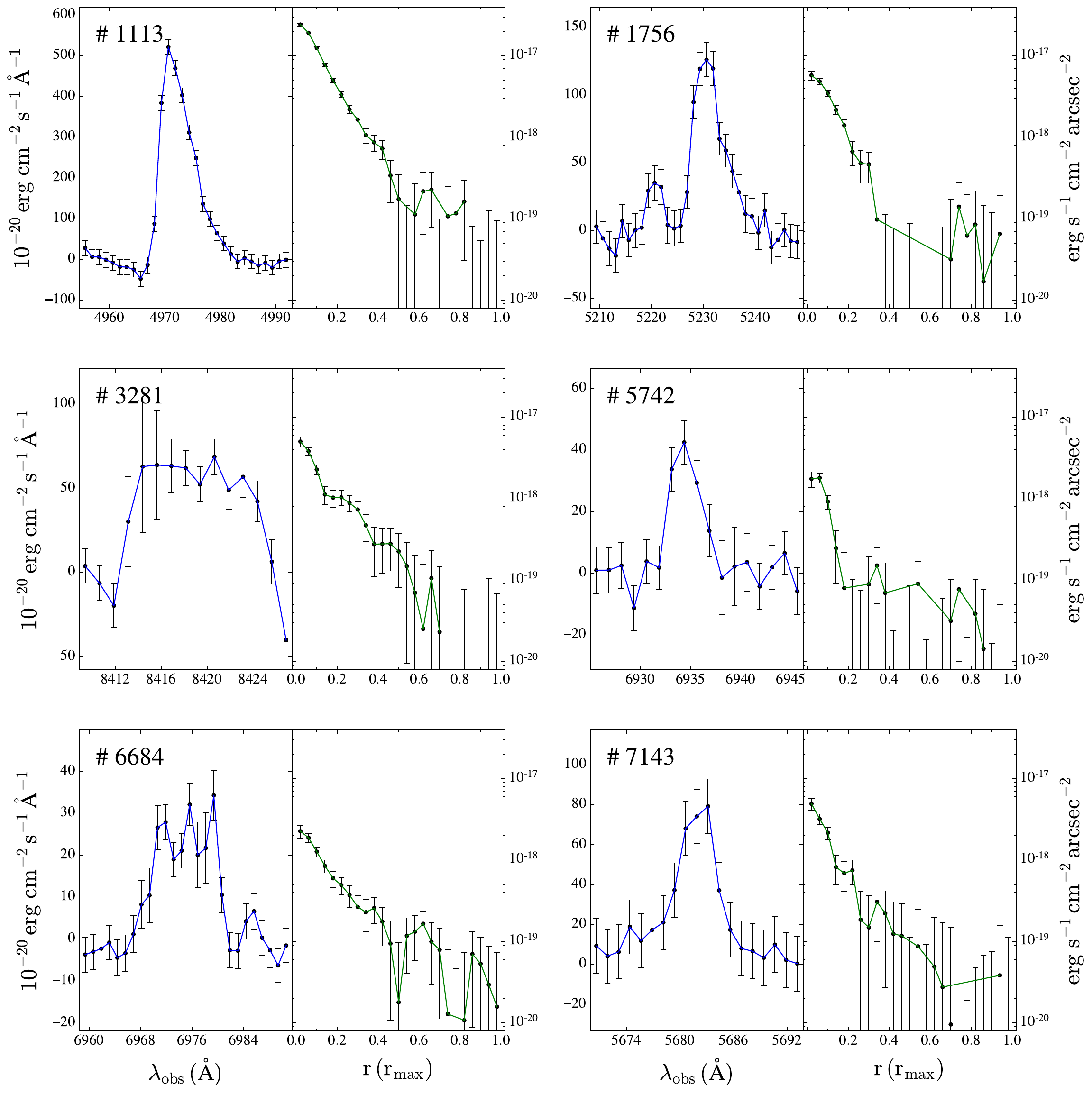}
    \caption{Examples of MUSE LAEs' Ly$\alpha$ spectra and surface brightness profiles (SBPs). The left of each panel presents the spectrum for each object, while the right shows the corresponding SBP. The MUSE ID for each object is indicated in the upper left corner of each panel. The spectra exhibit various shapes, such as a clear red peak, a double peak, or a broad emission line (see objects $\#$1113, $\#$1756, $\#$3281).}
    \label{fig:Ex_data_samples}
\end{figure}

Figure \ref{fig:Ex_data_samples} illustrates example spectra and SBPs, showing a range of data qualities and shapes of spectra and SBPs. As explained in the next section, the model adopted in \citet{Song2020} assumes an outflowing halo model and has the limitation of being able to reproduce only spectra that are red peak-dominated. Therefore, we proceeded with modeling only the 163 galaxies with red peak-dominated spectra.

\begin{figure}[t]
    \centering
    \includegraphics[width=10cm]{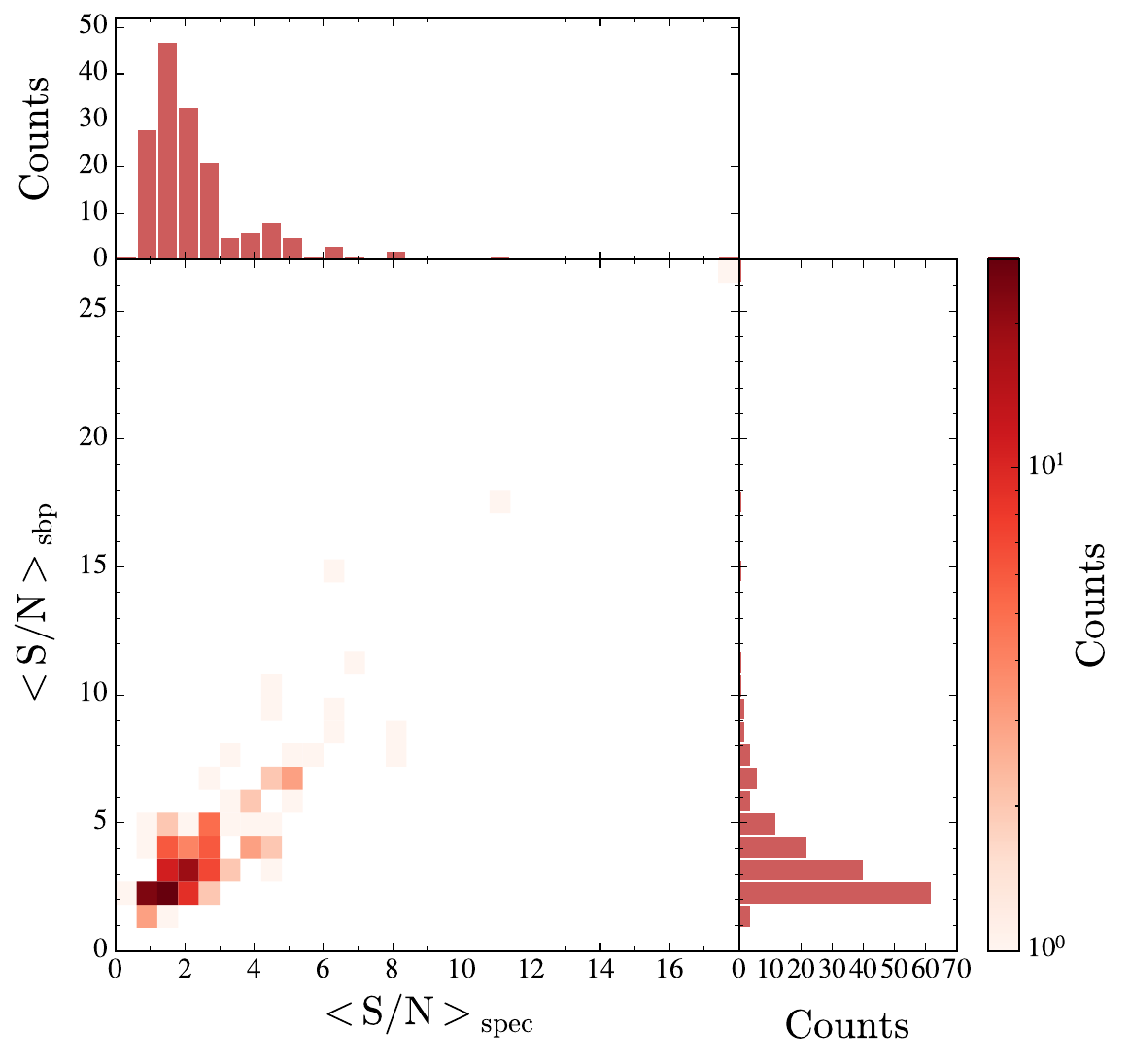}
    \caption{Distribution of S/N obtained from the spectrum ($\mathrm{\left<S/N\right>_{spec}}$) and the SBP ($\mathrm{\left<S/N\right>_{sbp}}$). The histograms in the upper left and lower right represent the individual distributions of $\mathrm{\left<S/N\right>}$ for the spectrum and SBP, respectively.}
    \label{fig:sn}
\end{figure}

While the MUSE IFU covers a wide wavelength range, we limited our analysis to a narrow range of rest-frame 6.13\AA\, around the spectral peak. The mean signal-to-noise ratio (S/N) of spectrum data points of each galaxy ranges from as low as 0.5 to as high as 17.8. For SBPs, we observed S/N ranging from 1.3 to 26.8.
Figure \ref{fig:sn} displays S/N distributions obtained from spectrum for the narrow range and the SBP, indicating that most objects have a marginal S/N between one and two.

\section{Modelled spectra and surface brightness profiles from Monte Carlo radiative transfer calculations for expanding halos}\label{sec:mod}

\citet{Song2020} performed Monte Carlo Radiative Transfer (MCRT) calculations for Ly$\alpha$ based on an expanding halo model described by six parameters. 
Table \ref{tab:T1} presents the six parameters along with their respective parameter spaces. 
While simple, this model provided good fits to the spectra and SBPs of eight MUSE galaxies in the previous study, and it is now being applied to a larger sample that is 20 times greater.
Here, we provide an overall description of the model. 

A halo is assumed to be spherically symmetric with an exponential HI density profile of a scale radius $rs_{\rm HI}$. 
The total amount of HI is parameterized by the optical depth $\tau_{0}$, which is defined as the integral from the center to the edge of the halo for a photon at the Ly$\alpha$ central wavelength, assuming a static medium. 
The medium temperature is fixed at $10^4$K. 
The bulk motion of the medium is assumed to be outflowing, leading to red-peak-dominated spectra, which are commonly observed in most star-forming galaxies.
The outflowing motion is parameterized by three parameters: $V_{\rm peak}$, $r_{\rm peak}$, and $\Delta V$, which represent the maximum outflow velocity, and the radius where $V_{\rm peak}$ occurs, and the velocity difference at $r_{\rm peak}$ and at the edge, respectively.
The velocity is modeled to vary linearly, increasing in the inner region and decreasing in the outer region due to feedback mechanisms and gravitational deceleration, respectively.

The initial spatial distribution of Ly$\alpha$ photons is assumed to follow that of the UV continuum: an exponential profile in the projected space and a modified Bessel function of the second kind in the three-dimensional space, since the main source of Ly$\alpha$ is the recombination line from \ion{H}{2} regions that are traced with the UV continuum.
Therefore, the initial distribution of Ly$\alpha$ photons is characterized by the UV continuum scale radius $rs_{\rm cont}$.
The initial frequency distribution is assumed to be a Voigt profile of temperature $10^4$K, a typical value for \ion{H}{2} regions.
It is worth mentioning that the models were run with uniform spatial and frequency distributions, which allows for the construction of any desired distributions by appropriately weighting each photon.

Redshift is also considered a parameter because it directly influences the shape of the spectrum; consequently, other parameters that affect the spectral shape cannot be accurately constrained if the redshift estimates are incorrect.
If no lines other than Ly$\alpha$ are available, the redshift is estimated based on the position of the Ly$\alpha$ peak, assuming it corresponds to the rest-frame wavelength of 1215.67\AA.
This estimation does not consider frequency diffusion caused by resonant scattering, which leads to an overestimation of the redshift.
This is likely the case for most of MUSE galaxies; therefore, redshift adjustments are made through modeling within the range of $z_{\rm MUSE}-0.015$ to $z_{\rm MUSE}+0.002$, where $z_{\rm MUSE}$ is the estimate provided by \citet{Leclercq2017}.
The adjustment window is broader than the one used by \citet{Song2020} to account for a variety of cases of the larger sample.

Dust-to-gas ratio (DGR), which was included as a parameter in the previous study, is not considered in this study. 
As seen in the posterior distributions in \citet{Song2020}, DGR is poorly constrained based on the shapes of the spectrum and SBP.
Its influence on the shapes remains subdominant until the optical depth reaches a significantly large value of $\log \tau_0\sim7.2$, corresponding to a hydrogen column density of $3\times 10^{20} {\rm cm^{-2}}$ \citep[see Figure C4 in][]{Song2020}.
Therefore, DGR was excluded from the model parameter set, resulting in a significant reduction in the computation time for the fitting procedure.

It is worth noting that the IGM plays a crucial role in shaping the Ly$\alpha$ spectra of LAEs at high redshifts, preferentially attenuating the blue side of the Ly$\alpha$ central wavelength and producing red peak-dominated spectra. Thus, while the red peak dominance observed in MUSE LAEs may reflect the ISM and CGM conditions, it could also result from the IGM attenuation. Notably, the mean IGM transmission decrease dramatically from approximately 60\% at $z=$3 to nearly zero at $z=$6 \citep{Inoue2014}, underscoring the importance of incorporating IGM transmission effects into Ly$\alpha$ radiative transfer modeling, especially for high-redshift galaxies. However, the significant object-to-object variation in IGM transmission, influenced by factors such as local velocity fields \citep[e.g.,][]{Park2021}, introduces additional uncertainties when applying the mean transmission. Moreover, the shape of Ly$\alpha$ spectrum emerging from the CGM remains uncertain. Therefore, to simplify our analysis, we disregard the effects of the IGM in this study.

\begin{deluxetable}{ll}
    \tablewidth{0pt}
    \caption{Parameters for an expanding halo model}
    \label{tab:T1}
    \tablehead{\colhead{Parameter} & \colhead{Values}}
    \startdata
    $rs_{\rm HI}\;(r_{\rm max})$   & $[0.1,0.2,0.3,0.4,0.5,0.6,0.7,0.8,0.9]$ \\
    $r_{\rm peak}\;(r_{\rm max})$  & $[0.0,0.1,0.2,0.3,0.4,0.5,0.6]$ \\
    $V_{\rm peak}\;(\mathrm{km\;s^{-1}})$  & $[100,200,300,400,500]$ \\
    $\Delta V\;(\mathrm{km\;s^{-1}})$      & $[-500,-450,-400,...,-100,-50,0]$ \\
    $\log\tau_{0}$ & $[5.7,6.0,6.3,6.6,6.9,7.2]$ \\
    $z$            & $z_{\rm MUSE}+[-0.015,-0.014,-0.013,...,+0.001,+0.002]$ \\
    \enddata
    \tablecomments{
        $rs_{\rm HI}$: The scale radius of the medium density distribution. It is normalized by the size of the domain, $r_{\rm max}$. \\
        $r_{\rm peak}$: The radius when the outflowing velocity of the medium reaches its peak, also normalized by $r_{\rm max}$. \\
        $V_{\rm peak}$: The peak velocity of the outflowing medium in the unit of $\mathrm{km\;s^{-1}}$. \\
        $\Delta V$: The difference between the velocities at $r_{\rm max}$ and $V_{\rm peak}$ in the unit of $\mathrm{km\;s^{-1}}$. \\
        $\log\tau_{0}$: The optical depth at the Ly$\alpha$ central wavelength. \\
        $z$: Redshift. $z_{\rm MUSE}$ is a rough estimate for the redshift of the system derived from observation.
    }
\end{deluxetable}

To find the best-fit model for each observed galaxy, the likelihood of each model is calculated by comparing its model spectrum and SBP to the observed ones.
The model spectrum and SBP are constructed following the method in \citet{Song2020}, taking into account the aperture size (for the construction of a spectrum), bandwidth (for the construction of a SBP), spectral resolution ($R\sim3000$), and point-spread function (PSF with a FWHM of $0.7\arcsec$) from the observation.

Since the aperture sizes and bandwidths, which vary among different galaxies, were not explicitly provided in \citet{Leclercq2017}, we aimed to estimate these values as closely as possible based on the descriptions in \citet{Leclercq2017}.
For the aperture size, it is determined by the size of HST segmentation mask for each object (i.e., the size of the rest-frame UV continuum emission) convolved with the MUSE PSF.
Since the UV continuum emission appears very compact ($\lesssim 0.13\arcsec$, which is $\lesssim1\,{\rm kpc}$ at $z=$3), the aperture size is essentially constrained by the MUSE PSF.
We tested whether the modeling results vary when the aperture size changes from $0.5\arcsec$ to $3\arcsec$.
Although some variation in the best-fit parameters was found for apertures larger than $1\arcsec$, the variation was marginal.
Moreover, the aperture sizes shown in Figures 2 and 3 of \citet{Leclercq2017} are $\lesssim1\arcsec$ independent of redshift and luminosity.
Therefore, we fixed the aperture size for the spectrum to $1\arcsec$ for all objects.
For the bandwidth, which was defined to maximize the S/N in \citet{Leclercq2017}, we determined a bandwidth that includes data points with small errors around the peak by visually inspecting the individual spectra.
This results in bandwidths ranging from 3.9 to 14.5\AA\, with a mean value of 8.1\AA\, \citep[c.f., it ranged from 2.5 to 20\AA\, with a mean value of 6.25\AA\, in][]{Leclercq2017}.
Although the bandwidths from this study and \citet{Leclercq2017} do not match exactly, our test on the impact of varying the bandwidth by 20\% and 40\% suggests that this difference is likely negligible.

As done in \citet{Song2020}, the likelihood of a model was calculated using the Neyman's chi-square statistic, assuming that the errors in the data follow a Gaussian distribution: 
\begin{equation}
\ln \mathcal{L} = -\frac{1}{2}\sum\limits_{i}\left(\frac{\mathcal{O}_{i}-\mathcal{M}_{i}}{\sigma(\mathcal{O}_{i})}\right)^2 , 
\end{equation}
where $\mathcal{M}$ and $\mathcal{O}$ denote the values of the model and the observation, respectively, and $\sigma(\mathcal{O})$ represents the observational error. 
Here, $i$ refers to the $i$-th wavelength bin in a spectrum or the radial bin in an SBP.
The total likelihood was computed by summing the likelihoods calculated from the spectrum and SBP, respectively, based on which the best-fit parameters were derived. 
Additionally, we determined and compared the best-fit parameters based on the individual likelihood distributions for the spectrum and SBP.
We primarily adopted Maximum Likelihood Estimation (MLE) to identify the best-fit parameters, and additionally evaluated Maximum A Posteriori (MAP) estimation under the assumption of uniform priors and compared the outcomes from both approaches.
While MLE selects the model with the highest likelihood as the best fit, MAP estimation determines the best-fit parameter based on the peak of its 1D marginal posterior distribution, obtained by marginalizing over the other parameters.

\section{Results and Discussion}\label{sec:res}
\subsection{Fitting assessment}\label{sec:fitting}

\begin{figure}[h!]
   \centering
   \includegraphics[width=15cm]{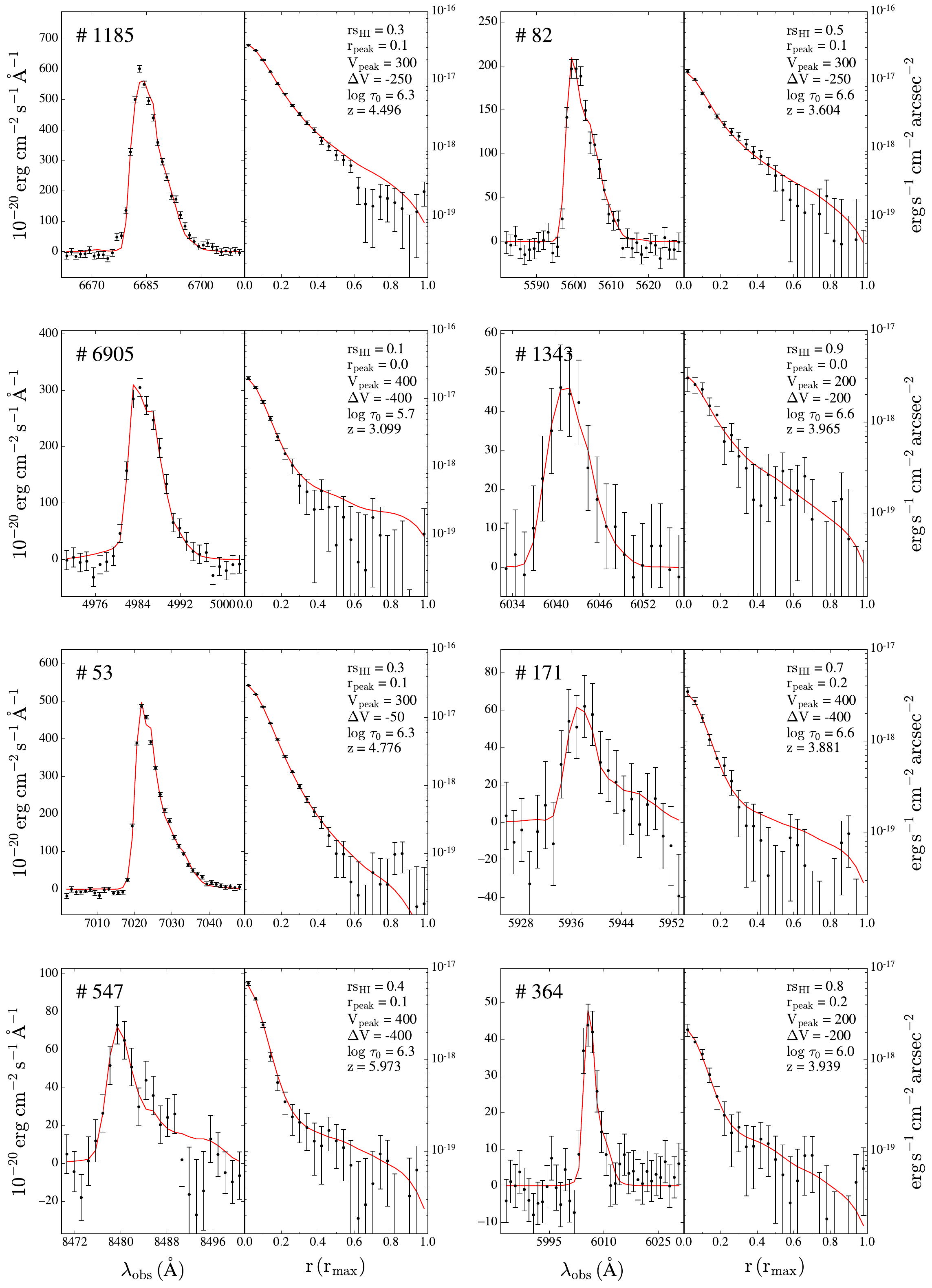}
   \caption{Fitting results for the eight objects analyzed in the previous study \citep{Song2020}. The left and right sides of each panel show the spectrum and SBP, respectively, with the observed data (black dots with error bars) and the best-fit model (red solid line). The MUSE ID is indicated in the upper left of each panel, and the best-fit parameter set used to construct the best-fit model is presented in the upper right.}
   \label{fig:Fitting_prof}
\end{figure}

\begin{figure}[h!]
   \centering
   \includegraphics[width=10cm]{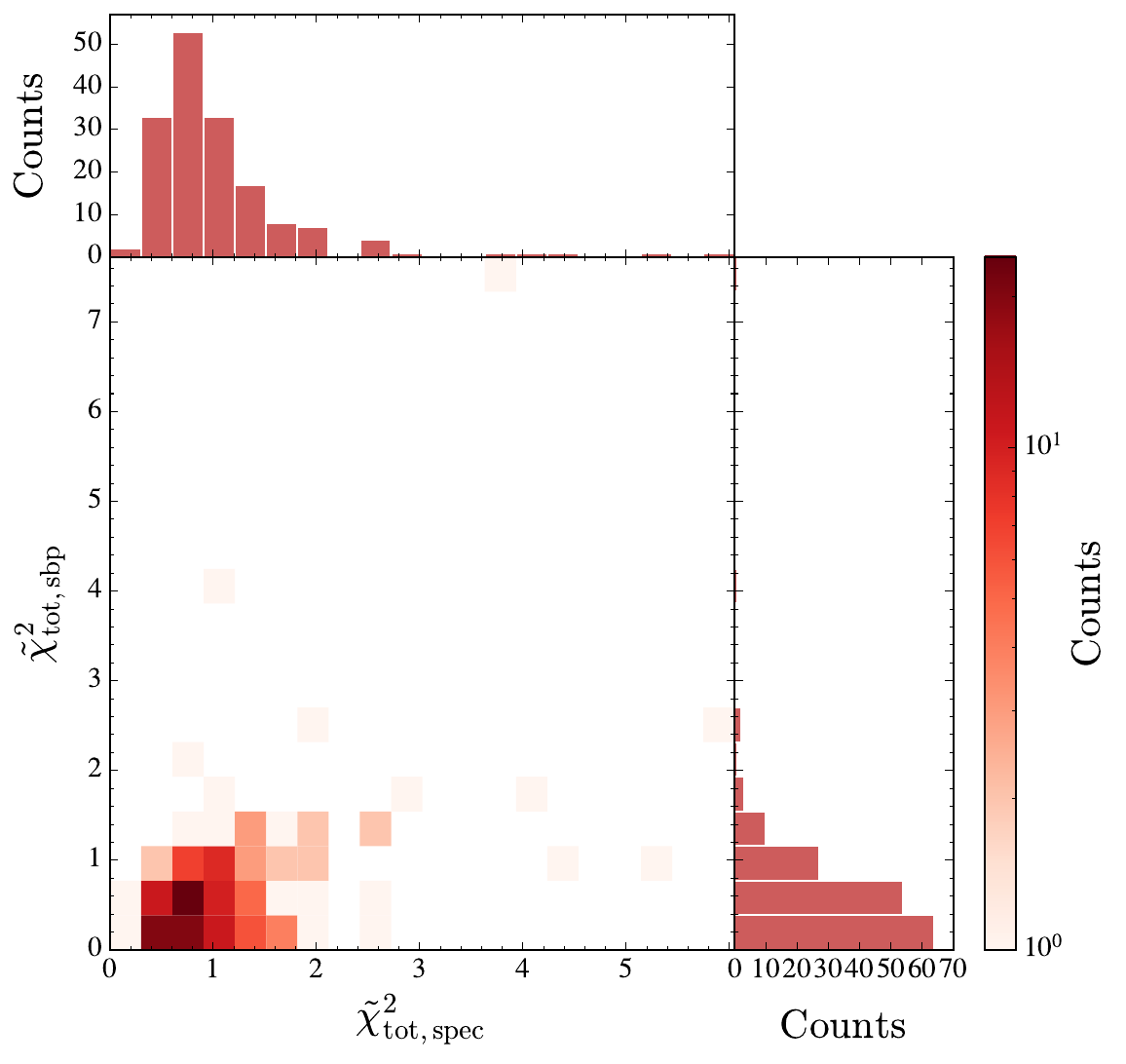}
   \caption{The reduced $\chi^2$ values for the 163 galaxies. $\tilde{\chi}_{\rm tot,spec}^2$ and $\tilde{\chi}_{\rm tot,sbp}^2$ represent the reduced $\chi^2$ values for the model spectra and the model SBP, respectively, based on the given best-fit parameter set.}
   \label{fig:Chi}
\end{figure}

We successfully determined the best-fit parameter sets for 163 individual galaxies, accurately reproducing their observed spectra and SBPs.
Figure \ref{fig:Fitting_prof} displays the observed and modeled spectra and SBPs for the eight galaxies analyzed in \citet{Song2020}.
Figure \ref{fig:Chi} presents the overall fitting results, showing the distribution of reduced $\chi^2$ values calculated separately for spectra and SBPs using the final best-fit parameter set.
For most galaxies, the reduced $\chi^2$ values are close to one, confirming the success of the fit for both spectra and SBPs.
Only {\bf a small number of objects} exhibit large reduced $\chi^2$ values, primarily due to exceptionally small observational errors in certain data points.
Nevertheless, even in these cases, the overall shapes of the spectra and SBPs are reasonably well reproduced.

\begin{figure}[t]
   \centering
   \includegraphics[width=\textwidth]{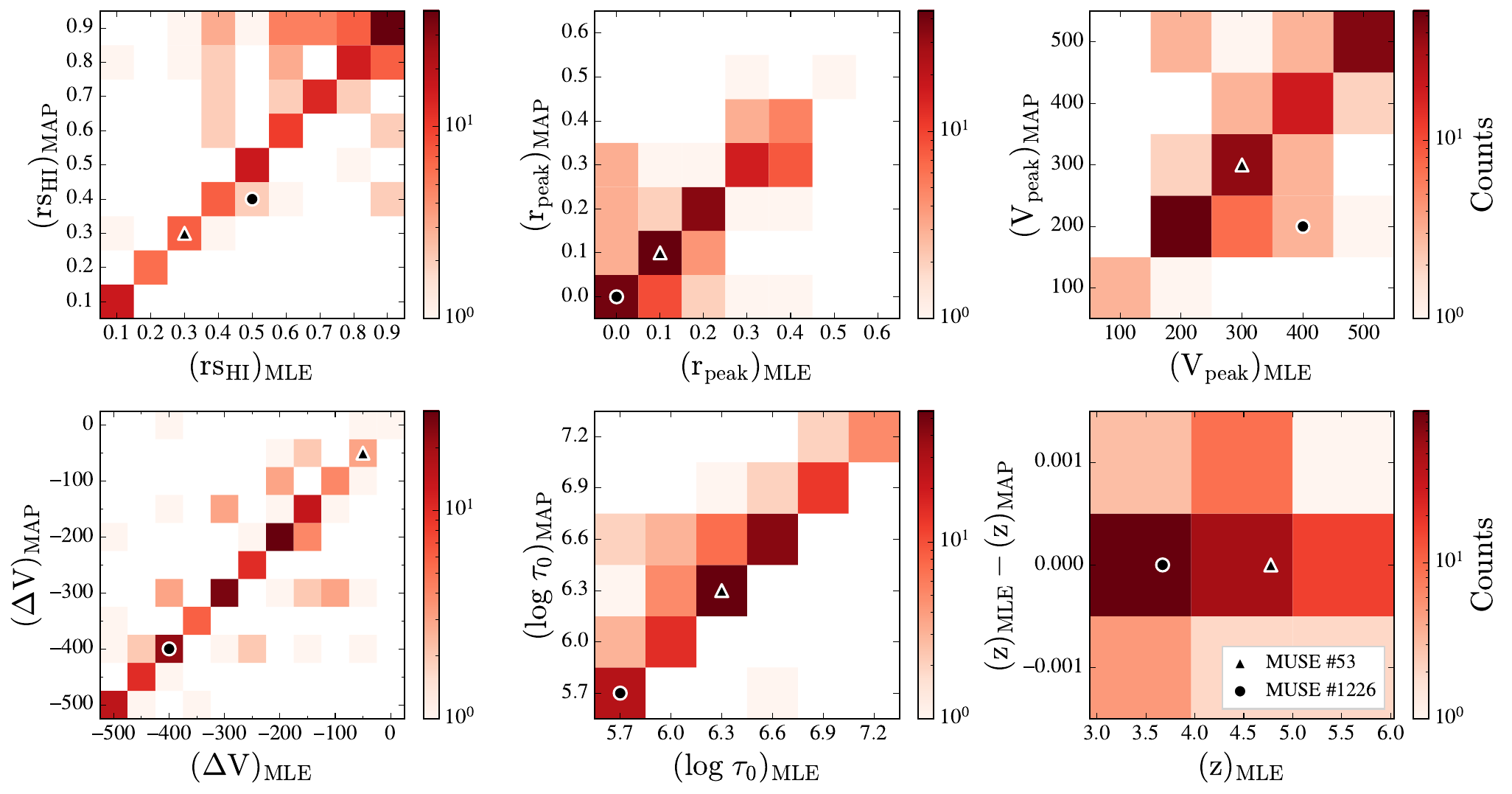}
   \caption{Comparison of the best-fit parameters obtained with and without marginalization. Each panel represents the two-dimensional histogram for one of our model parameters, obtained from all samples, illustrating the best-fit parameters obtained after marginalization (denoted by the subscript ``MAP'') compared to those derived solely from likelihood values (``MLE''). The triangles and circles denote the object with high S/N and the one with low S/N, respectively.}
   \label{fig:Best_marg}
\end{figure}

Since the best-fit parameters can vary depending on whether MLE or MAP is applied, we compared the results of both approaches.
Figure \ref{fig:Best_marg} displays the best-fit parameters obtained from each method for all objects, along with the results for two specific objects--one with high S/N and one with low S/N--indicated by triangles and circles, respectively.
The two methods generally yield consistent results with each other, suggesting that the posterior distributions are not peculiar, the uniform priors are appropriate, and the data is sufficiently informative for most of our sample galaxies.
Albeit weak, there is a tendency for the results of the two methods to disagree when the S/N of the observed data is low, as indicated by the circles. While MAP is expected to provide more stable constraints on the best-fit parameters under low S/N conditions,  it also heavily depends on the choice of prior. Therefore, we chose to use MLE for determining the best-fit parameters as it offers direct results based solely on the data, without being influenced by potentially misleading priors.
Ultimately, since MLE and MAP generally yield similar results, we anticipate that the choice of method will not significantly influence the conclusions of this study.

Finally, we validated our fitting results by comparing the results for the eight galaxies analyzed in \citet{Song2020} with those obtained in this study.
The best-fit parameters for the eight galaxies are summarized in Table \ref{tab:T2}, and the resulting spectra and surface brightness profiles are presented in Figure \ref{fig:Fitting_prof}.
The best-fit parameter values for MUSE \#171, 547, and 6905 differ noticeably from those reported in \citet{Song2020}.
Several factors may contribute to the discrepancies.
First, unlike the previous study, which manually extracted observational data from the figures in \citet{Leclercq2017}, potentially leading to inaccuracies in the extracted values (especially an overestimation of error values), we obtained the data directly through the MUSE collaboration.
Second, to improve the fit of the SBP over a large dynamic range of the surface brightness, the likelihood was calculated on a logarithmic scale by taking the logarithm of both the observed and model values.
Lastly, while the previous study employed the MAP method, we utilized the MLE method, which may also account for some discrepancies, albeit likely to a minor extent.

\begin{figure}[t]
   \centering
   \includegraphics[width=\textwidth]{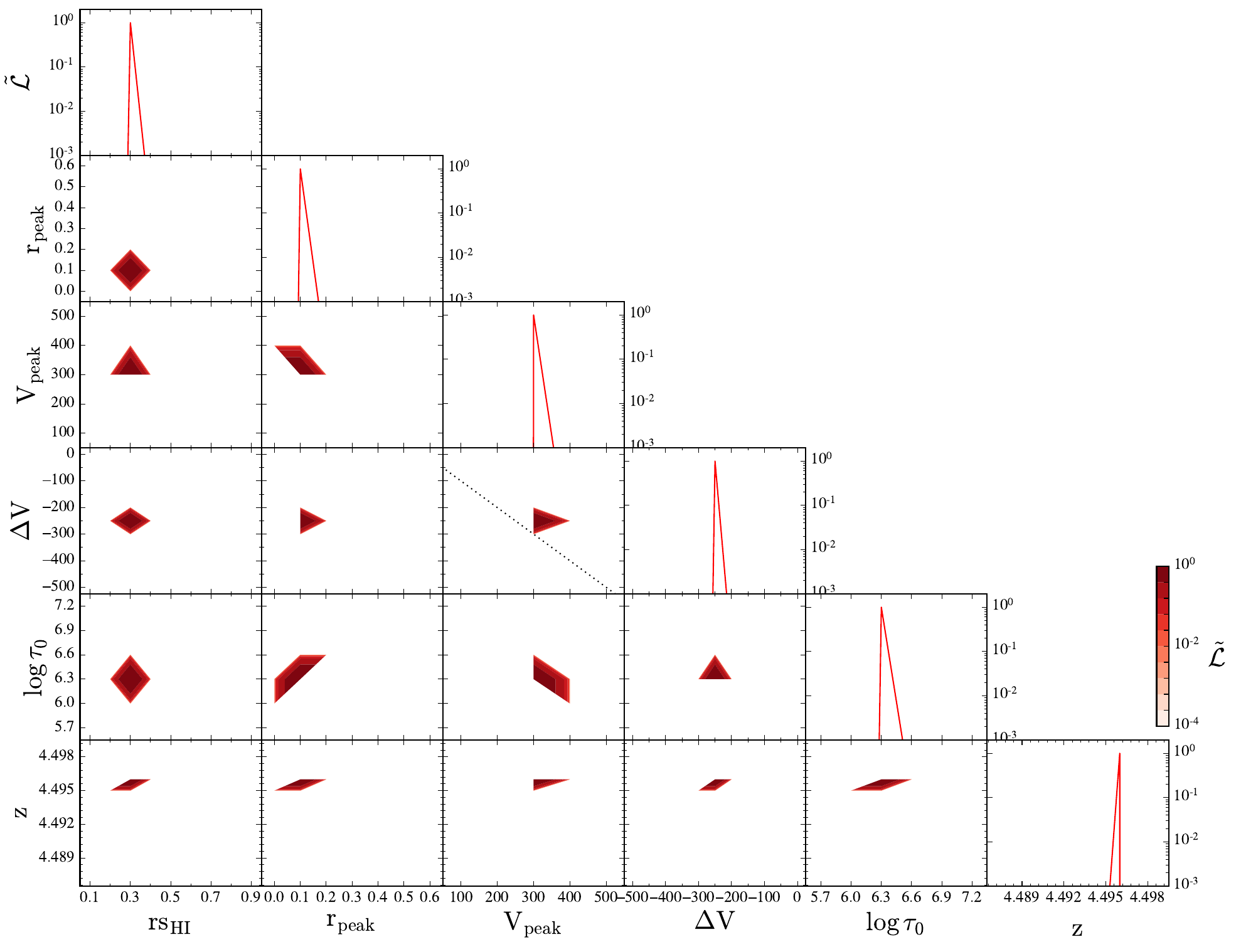}
   \caption{An example of the marginal posterior distribution. The top panel in each column represents the 1D posterior distribution, while the remaining panels show the 2D posterior map for the total likelihood of object $\#$1185. The tilde denotes that the posterior values have been normalized to their maximum, and the dotted line indicates the allowed $\Delta V$ range.}
   \label{fig:Pos1}
\end{figure}

\begin{figure}[t]
   \centering
   \includegraphics[width=\textwidth]{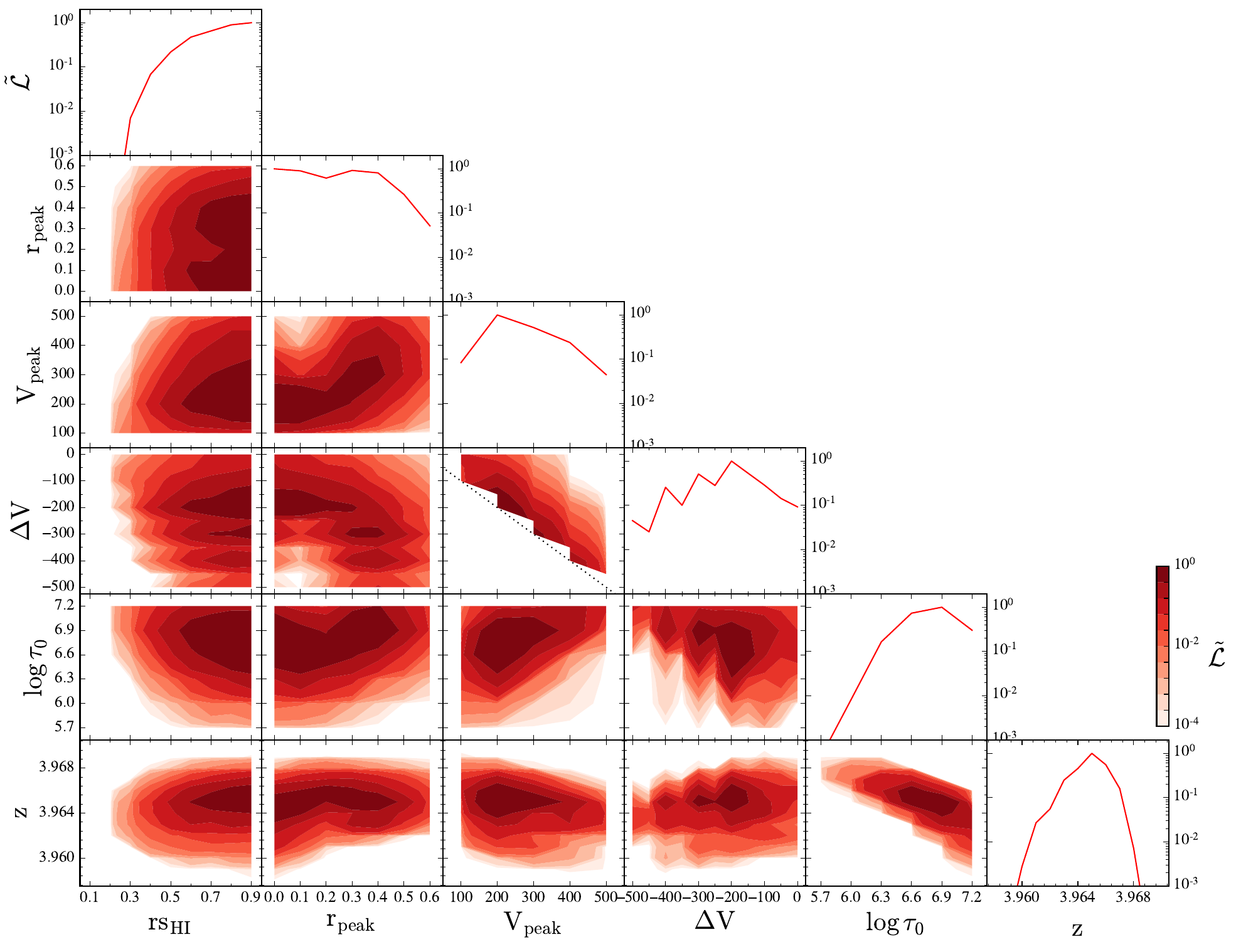}
   \caption{Similar to Figure \ref{fig:Pos1}, but for object $\#$1343}
   \label{fig:Pos2}
\end{figure}

Since the present results were obtained using actual observational data rather than digitized data from figures, it is reasonable to adopt the fitting results from this study.
Additionally, the SBP likelihood calculated on a logarithmic scale provides more comprehensive fits across the full range of SBPs, as demonstrated by the comparison between our fitting results and the previous ones, particularly for MUSE \#1343 \citep[see Figure \ref{fig:Fitting_prof} and Figure B12 in][]{Song2020}.
A comparison of the marginal posterior distributions showed that our posterior distributions are sharper than those in the previous study \citep[Figures \ref{fig:Pos1} and \ref{fig:Pos2} versus Figures 5 and B3 in][]{Song2020}, mainly due to smaller errors in the data used in this study.
With the sharper posterior distributions, the degeneracies between the parameters observed in the previous study are more clearly visible in the present study (e.g., $z$--$\tau_0$, $r_{\rm peak}$--$V_{\rm peak}$, and $\tau_{0}$--$V_{\rm peak}$).

Overall, the fitting results appear reliable, enabling further investigations with the large sample of MUSE galaxies.
In Section \ref{sec:fitting2}, we examine the impact of using both spectra and SBPs on the determination of the best-fit parameters.
In Section \ref{sec:correlations}, we explore which parameter has the greatest influence on the spatial extent and spectral characteristics of Ly$\alpha$ emission in star-forming galaxies.

\begin{deluxetable*}{cccccccc}
\tablecaption{Comparison of the best-fit parameters of \citet{Song2020} and this study for the eight galaxies that were modeled in \citet{Song2020} \label{tab:T2}}
\tablewidth{0pt}
\tablehead{
\colhead{ID \#} & \colhead{Paper} & \colhead{$rs_{\rm HI}$} & \colhead{$r_{\rm peak}$} & \colhead{$V_{\rm peak}$} & \colhead{$\Delta V$} & \colhead{$\log \tau_0$} & \colhead{$z$}
}
\startdata
\multirow{2}{*}{1185} 
 & Song et al.   & 0.3  & 0.2  & 300 & -150 & 6.6 & 4.495 \\
 & this study    & 0.3  & 0.1  & 300 & -250 & 6.3 & 4.496 \\
\hline
\multirow{2}{*}{82}
 & Song et al.   & 0.5  & 0.1  & 300 & -250 & 6.6 & 3.604 \\
 & this study    & 0.5  & 0.1  & 300 & -250 & 6.6 & 3.604 \\
\hline
\multirow{2}{*}{6905}
 & Song et al.   & 0.1  & 0.0  & 300 & -300 & 6.3 & 3.098 \\
 & this study    & 0.1  & 0.0  & 400 & -400 & 5.7 & 3.099 \\
\hline
\multirow{2}{*}{1343}
 & Song et al.   & 0.8  & 0.4  & 200 & -200 & 6.9 & 3.964 \\
 & this study    & 0.9  & 0.0  & 200 & -200 & 6.6 & 3.965 \\
\hline
\multirow{2}{*}{53}
 & Song et al.   & 0.4  & 0.1  & 300 & -50  & 6.3 & 4.776 \\
 & this study    & 0.3  & 0.1  & 300 & -50  & 6.3 & 4.776 \\
\hline
\multirow{2}{*}{171}
 & Song et al.   & 0.8  & 0.0  & 200 & 0    & 6.3 & 3.882 \\
 & this study    & 0.7  & 0.2  & 400 & -400 & 6.6 & 3.881 \\
\hline
\multirow{2}{*}{547}
 & Song et al.   & 0.7  & 0.1  & 300 & -200 & 6.3 & 5.974 \\
 & this study    & 0.4  & 0.1  & 400 & -400 & 6.3 & 5.973 \\
\hline
\multirow{2}{*}{364}
 & Song et al.   & 0.9  & 0.3  & 200 & -200 & 6.0 & 3.939 \\
 & this study    & 0.8  & 0.2  & 200 & -200 & 6.0 & 3.939 \\
\enddata
\end{deluxetable*}

\newpage
\subsection{Simultaneous fitting of spectrum and surface brightness profile}
\label{sec:fitting2}
\begin{figure}[t]
   \centering
   \includegraphics[width=\textwidth]{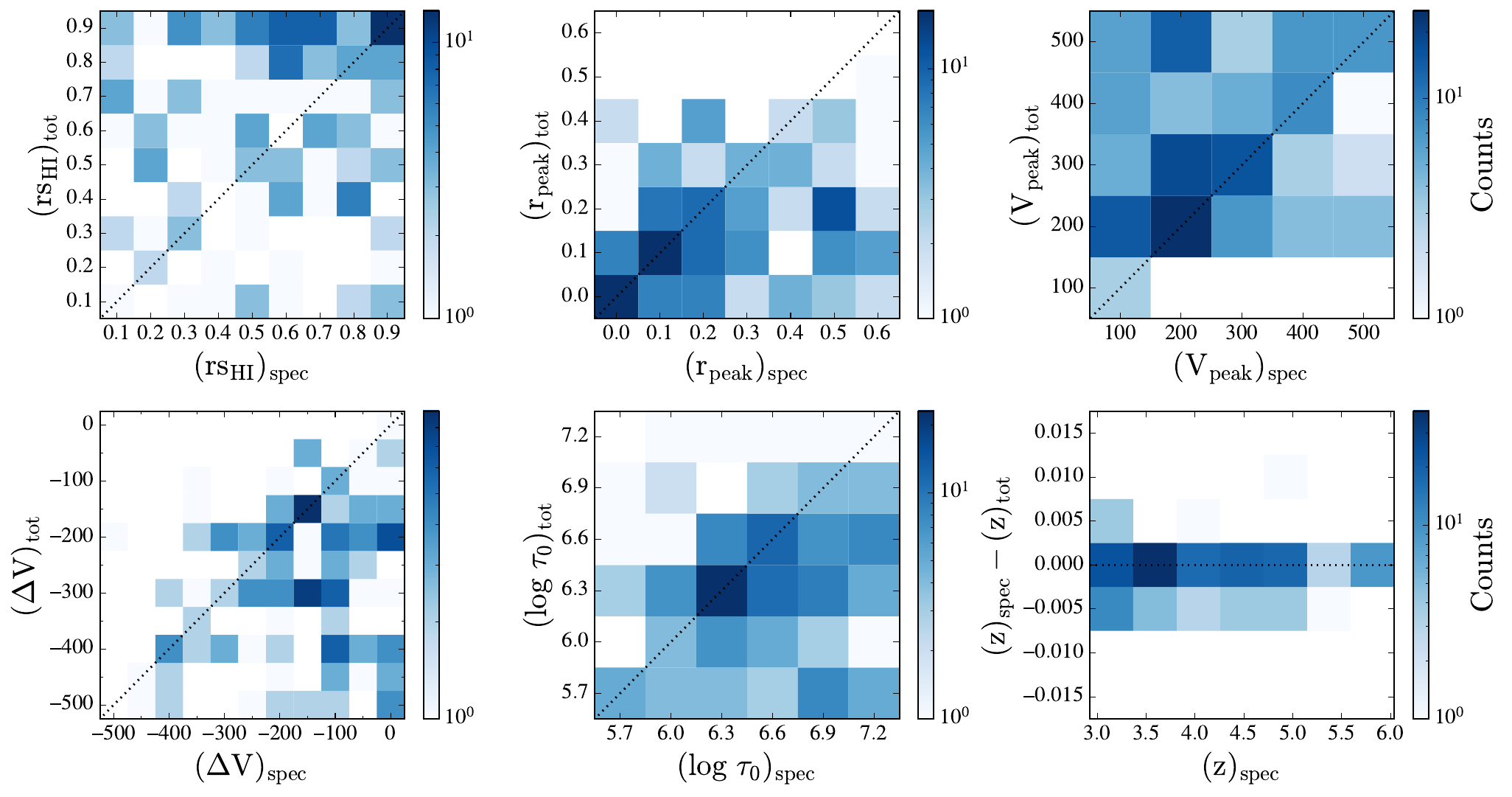}
   \caption{Comparison between best-fit parameters. The subscript ``spec'' indicates the best-fit parameter obtained by considering only the spectrum, while ``tot'' denotes the total best-fit parameter determined by considering both the spectrum and the SBP. The black dotted line represents $y=x$ line.}
   \label{fig:Best_spto}
\end{figure}

\begin{figure}[t]
   \centering
   \includegraphics[width=\textwidth]{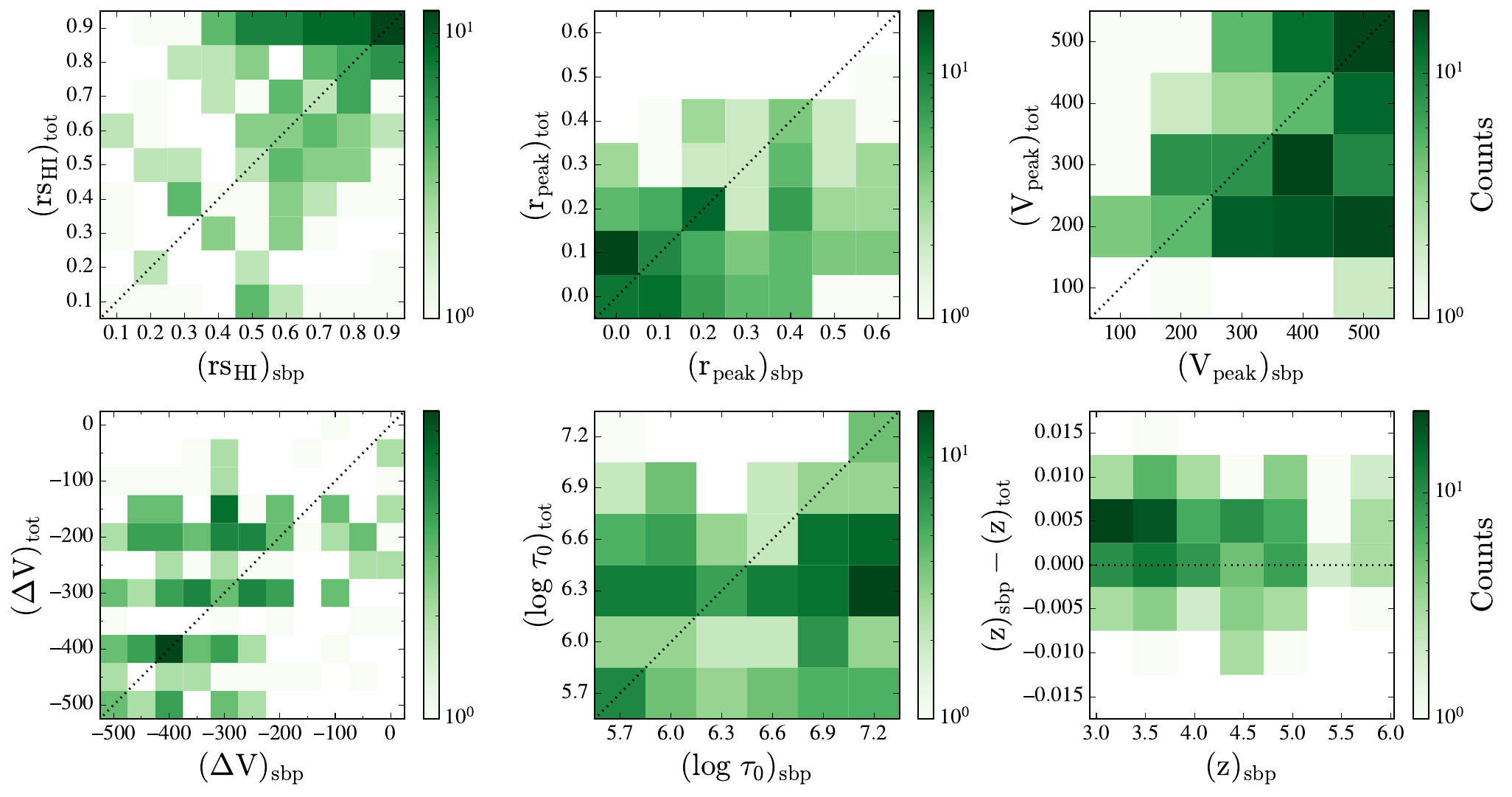}
   \caption{Similar to Figure \ref{fig:Best_spto}, but comparing the total best-fit parameters with those obtained by considering only the SBP.}
   \label{fig:Best_sbto}
\end{figure}

\citet{Song2020} demonstrated that simultaneously utilizing both spectra and SBPs helps break or reduce degeneracies between parameters, thereby tightening parameter constraints.
Here, we compare the best-fit parameters obtained based solely on either spectra (denoted with the subscript ``spec'') or SBPs (``sbp'') with those derived considering both quantities (i.e., based on the total likelihood distributions; ``tot'') in Figures \ref{fig:Best_spto} and \ref{fig:Best_sbto}.
Through this comparison, we investigated which--between the spectrum and the SBP--plays a key role in determining each parameter.

It is evident that $r_{\rm peak}$, $V_{\rm peak}$, $\tau_0$, and redshift are primarily determined by the spectrum, as indicated by the fact that most galaxies, represented by dark pixels, lie along the diagonal one-to-one correspondence line (or the horizontal $(z)_{\rm sbp}-(z)_{\rm tot}=0$ line in the case of redshift) in Figure \ref{fig:Best_spto}, while they are found at off-diagonal locations in Figure \ref{fig:Best_sbto}.
Comparison of the figures also reveals that $(\Delta V)_{\rm tot}$ values tend to align more closely with $(\Delta V)_{\rm sbp}$ than with $(\Delta V)_{\rm spec}$.
While the spectrum-only and SBP-only fittings generally predict large values for $rs_{\rm HI}$, the overall distribution suggests that the SBP has a greater influence on determining $rs_{\rm HI}$ than the spectrum; there are cases where the spectrum-only fitting (Figure \ref{fig:Best_spto}) yields significantly different predictions for $rs_{\rm HI}$ compared to the results obtained when the spectrum and SBP are simultaneously considered.

\begin{figure}[t]
   \centering
   \includegraphics[width=16cm]{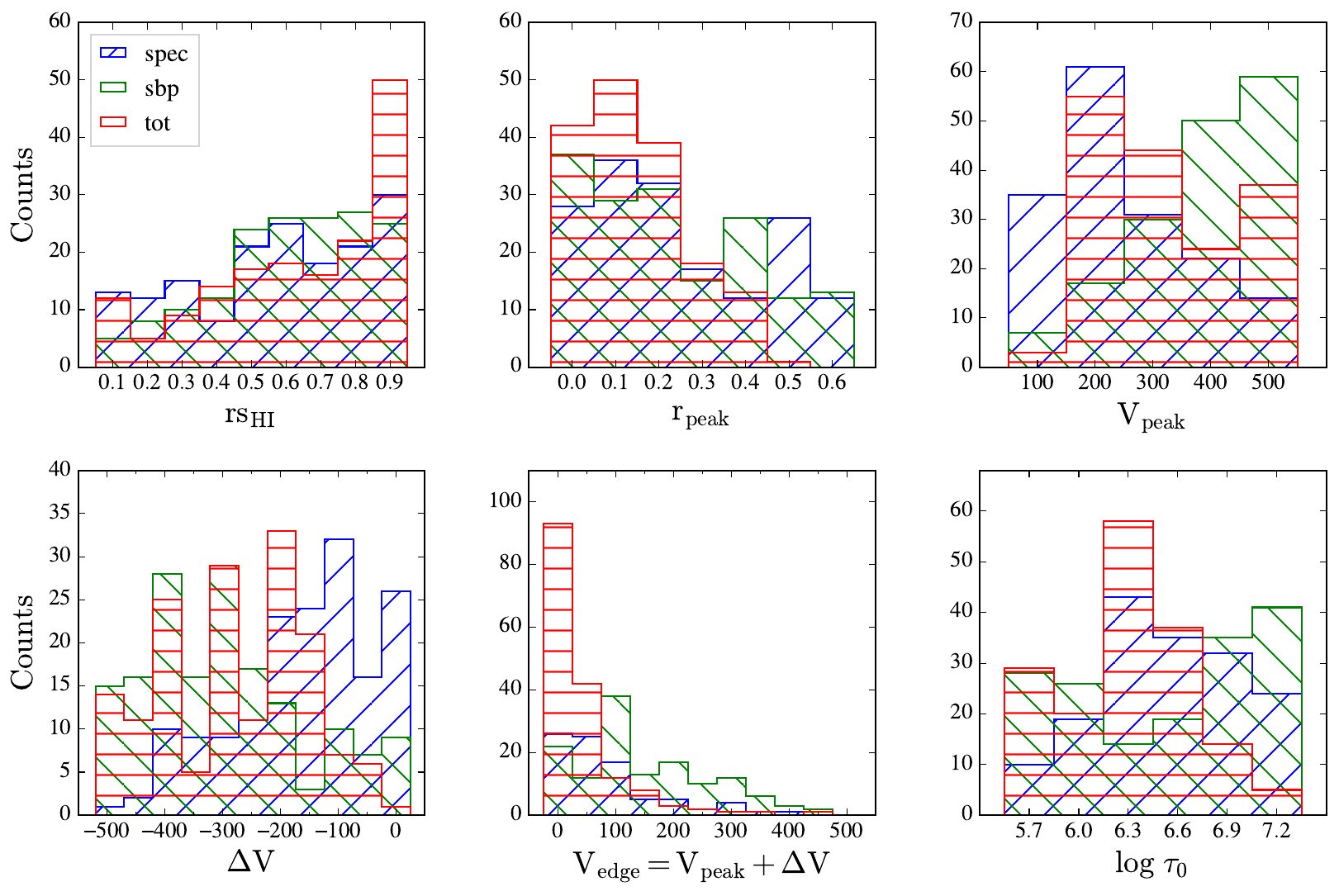}
   \caption{The histogram of best-fit values for each parameter. The blue area shows the distribution of best-fit parameters derived by considering only the spectrum, the green area illustrates the distribution based on only SBP, and the red area depicts the total best-fit parameter distribution when both the spectrum and SBP are considered.}
   \label{fig:Best}
\end{figure}

The large scatter in the overall distributions in Figures \ref{fig:Best_spto} and \ref{fig:Best_sbto}, despite a subset of galaxies where model parameters can be determined based solely on the spectrum, highlights the advantage of modeling both the spectrum and SBP simultaneously.
Figure \ref{fig:Best} shows the distributions of the best-fit values for each parameter when considering only the spectrum (blue hatched histogram), only the SBP (green), and both the spectrum and SBP (red).
These three distributions appear distinct, and each model parameter spans the full parameter range. However, the ``tot'' best-fit values of $rs_{\rm HI}$ and $V_{\rm edge}$ are noticeably concentrated at 0.9 $r_{\rm max}$ and 0 km s$^{-1}$, respectively.
These tendencies in $rs_{\rm HI}$ and $V_{\rm edge}$ suggest that the medium is likely to be largely extended and static at large radii.

\subsection{Correlations between observables and model parameters}
\label{sec:correlations}

\begin{figure}[t]
   \centering
   \includegraphics[width=\textwidth]{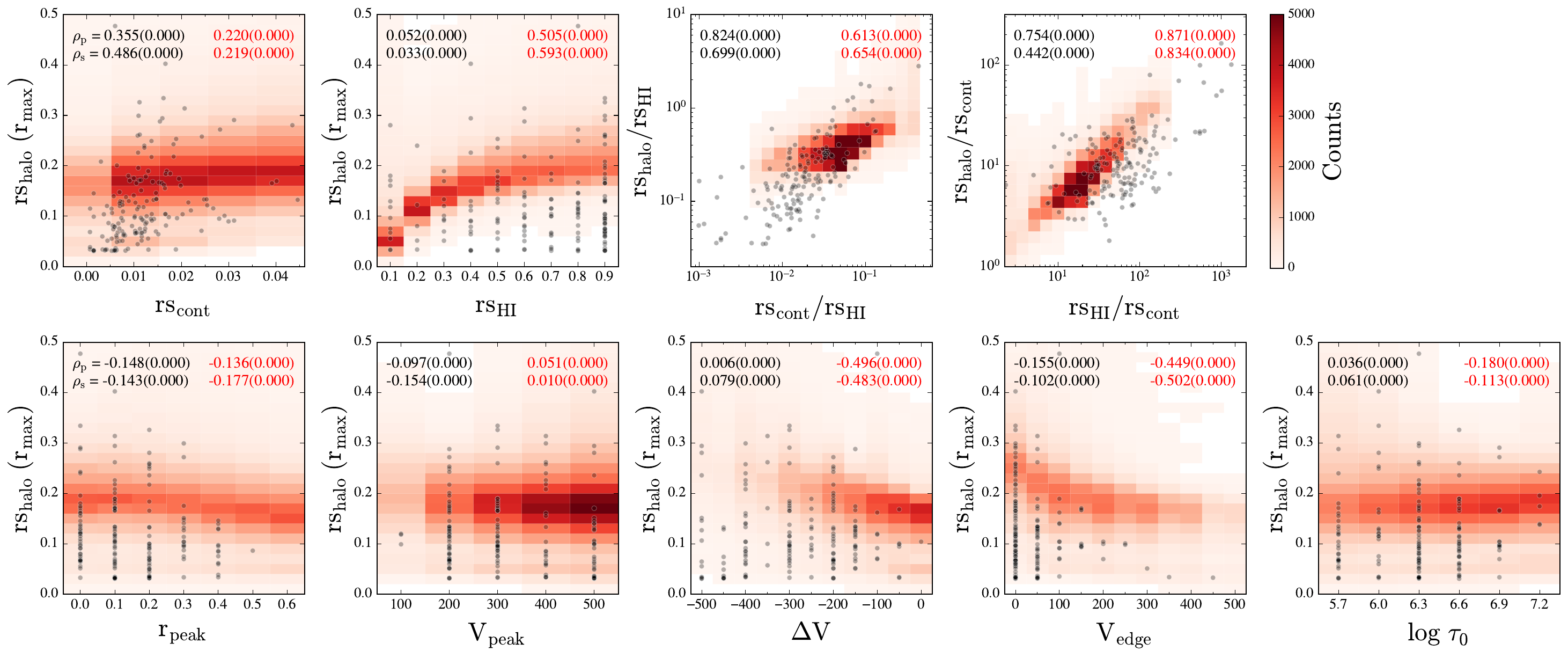}
   \caption{Correlation between the size of Ly$\alpha$ halo ($rs_{\rm halo}$, normalized by $r_{\rm max}$) and each of the model parameters.
   The distribution of the full set of models is shown as a 2D histogram in red and the best-fit models for observed galaxies are shown as gray dots. Since transparency is applied, overlapping dots appear darker. The Pearson correlation coefficient ($\rho_{\rm p}$) and Spearman rank correlation coefficient ($\rho_{\rm s}$) are shown in red for all models and in black for the best-fit models, with the corresponding $p$-value in parentheses.}
   \label{fig:rshalo}
\end{figure}

In this section, we explore the correlations between Ly$\alpha$ observables and best-fit model parameters (``tot'').
Figure \ref{fig:rshalo} shows the correlation between the size of Ly$\alpha$ halo and each model parameter, including the scale radius of UV continuum emission ($rs_{\rm cont}$), which is a measured quantity in observations.
The halo size in the model is defined based on a two-component exponential profile, following the procedure described in \citet{Leclercq2017}. The scale radius of the core component is set by the UV continuum emission.
Each correlation is explored using the full set of models (2D histogram in red) and the best-fit models for the MUSE galaxies (gray dots).
To quantitatively assess these correlations, we calculated the Pearson correlation coefficient ($\rho_{\rm p}$) and the Spearman rank correlation coefficient ($\rho_{\rm s}$), both shown in each panel along with their corresponding $p$-values.
The Spearman rank correlation coefficient is less sensitive to outliers, unaffected by scale, and applicable to both linear and non-linear correlations, providing additional insights complementary to those from the Pearson correlation coefficient.
We considered a correlation to be robust when the $p$-values for these two correlation coefficients were $\lesssim 0.05$.

As shown in \citet{Leclercq2017} (see their Figure 13) and illustrated again in Figure \ref{fig:rshalo}, \textbf{a moderate positive correlation between $rs_{\rm halo}$ and $rs_{\rm cont}$ is observed both in the full model set and in the best-fit subset}, which aligns with what we would expect for these two quantities.
Similarly, a strong correlation between $rs_{\rm halo}$ and $rs_{\rm HI}$ is expected and is indeed evident in the distribution of the full model set.
However, when only the best-fit models are considered, this correlation appears insignificant, primarily due to a sample bias toward models with an extended medium distribution (see Figure \ref{fig:Best}).
This highlights that the correlations observed in the best-fit models do not necessarily reflect those of the full model set.
Rather, they should be interpreted as characteristics of the specific sample, not as general properties of Ly$\alpha$ radiative transfer.
This discrepancy between the full model set and the best-fit subset not only reflects the sample bias but also highlights the intrinsically complex nature of Ly$\alpha$ radiative transfer.
In such systems, multiple physical parameters interact in nonlinear and degenerate ways to shape the emergent observables.
As a result, even moderate sample biases can lead to significant deviations in the apparent correlations, depending on which regions of parameter space are occupied.

The sample bias that weakens the $rs_{\rm halo}$--$rs_{\rm HI}$ correlation can be partially alleviated by considering the normalized quantity, $rs_{\rm HI}/rs_{\rm cont}$.
We also examine the correlation between $rs_{\rm halo}$ and $rs_{\rm cont}$ by normalizing the scales by $rs_{\rm HI}$.
With this normalization, the $rs_{\rm halo}$--$rs_{\rm HI}$ correlation re-emerges among the best-fit models (as shown in the upper fourth panel), although its strength is not fully recovered.
The $rs_{\rm halo}$--$rs_{\rm cont}$ correlation also becomes stronger when normalized (upper third panel).
Overall, the fact that these correlations remain among the strongest in both the full and best-fit model sets suggests that the spatial extents of the medium and the sources are the primary determinants of Ly$\alpha$ halo size.

In addition to the distributions of the medium and the source, the velocity structure of the medium also plays a key role in determining the extent of Ly$\alpha$ halos.
For Ly$\alpha$ halos to appear more extended, photons must scatter at larger radii, which requires a lower edge velocity ($V_{\rm edge}$).
This leads to an expected anti-correlation between $rs_{\rm halo}$ and $V_{\rm edge}$ (or $\Delta V$).
While this anti-correlation is clearly seen in the full model set, it is significantly diminished in the best-fit subset, likely due, once again, to sample bias arising from the preferential selection of models with low $V_{\rm edge}$.

Another noticeable anti-correlation exists between $rs_{\rm halo}$ and $r_{\rm peak}$, although it is relatively weaker than those discussed earlier.
This anti-correlation arises from the trend that the medium velocity at a given inner radius ($r < r_{\rm peak}$) tends to be lower when $r_{\rm peak}$ is larger.
Lower velocities at small radii increases the likelihood that Ly$\alpha$ photons will be scattered and trapped closer to the center, resulting in a more compact Ly$\alpha$ halo.
This anti-correlation is also present in the best-fit models with comparable strength, likely because the sample bias in $r_{\rm peak}$ is less pronounced than that in $rs_{\rm HI}$ or $V_{\rm edge}$.
It is worth noting that this anti-correlation would be weaker in models with high $V_{\rm peak}$, as the overall scattering probability would already be low and thus not be significantly enhanced by increasing $r_{\rm peak}$.

While strong correlations of $rs_{\rm halo}$ with $rs_{\rm cont}$, $rs_{\rm HI}$ and $V_{\rm edge}$ are evident, the lack of a strong correlation with $\tau_0$ is interesting.
This suggests that although a certain amount of material may be necessary for the formation of extended Ly$\alpha$ halos, their spatial extent is more strongly governed by the density and velocity distributions of the medium.

\begin{figure}[t]
   \centering
   \includegraphics[width=\textwidth]{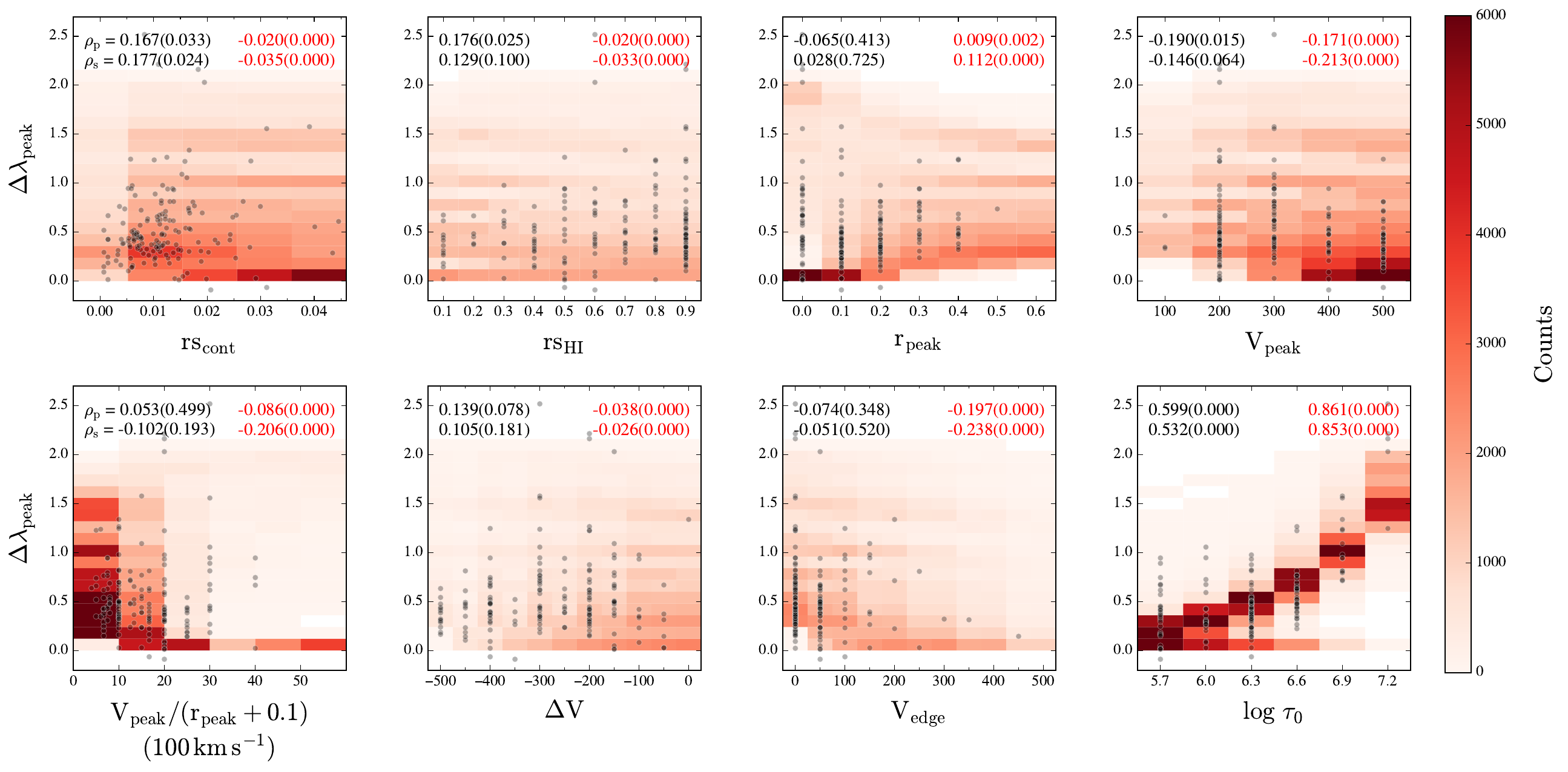}
   \caption{Similar to Figure \ref{fig:rshalo}, but for peak shift ($\Delta \lambda_{\rm peak}$).}
   \label{fig:peakshift}
\end{figure}

\begin{figure}[t]
   \centering
   \includegraphics[width=\textwidth]{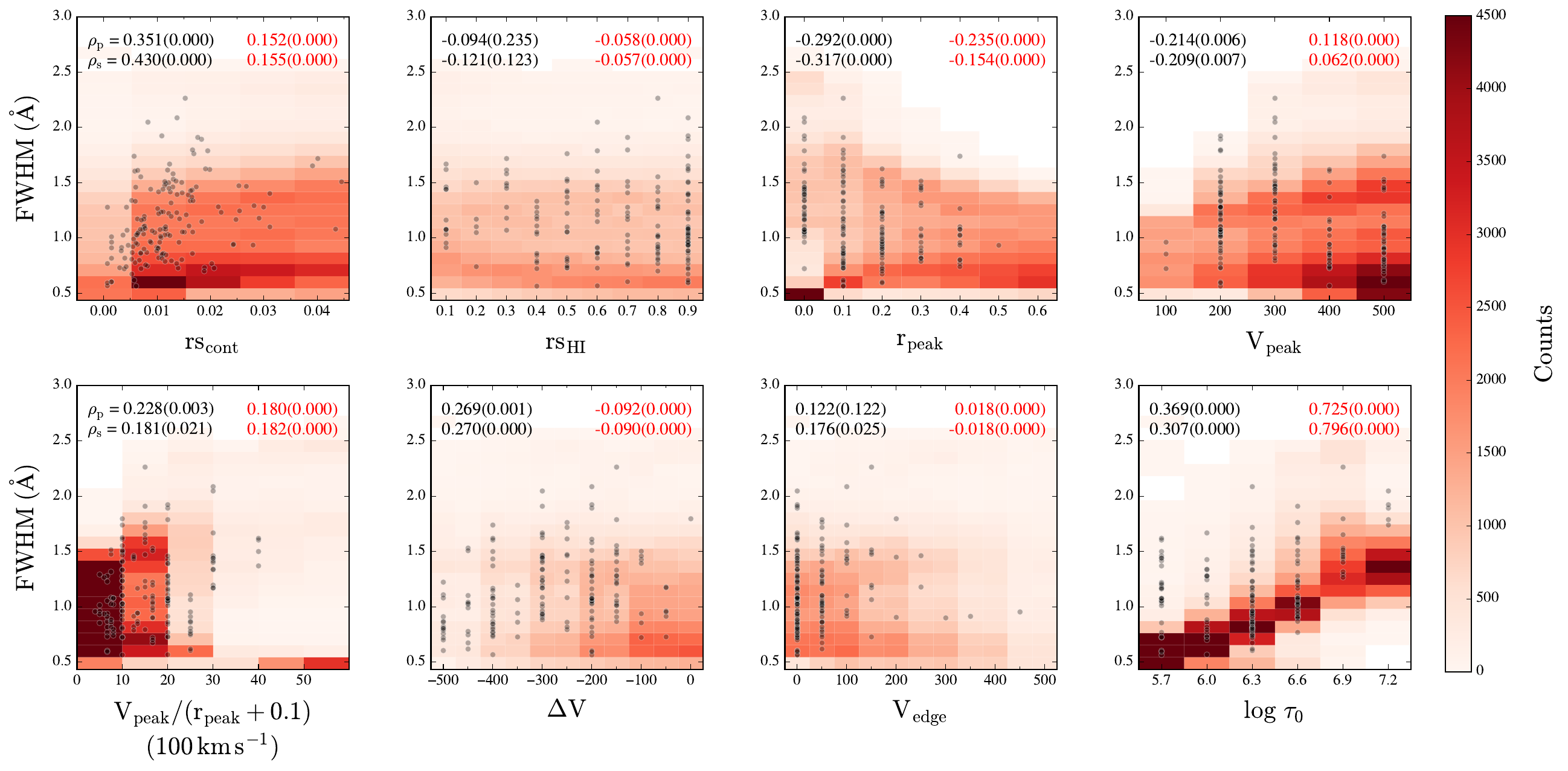}
   \caption{Similar to Figure \ref{fig:rshalo}, but for the FWHM of the spectrum.}
   \label{fig:FWHM}
\end{figure}

The dominant parameters determining the peak shift or FWHM were previously analyzed in \citet{Song2020} using the full model set. 
Their analysis revealed a clear positive correlation with optical depth and suggested potential correlations with parameters related to the medium's velocity profile (see their Figure 11).
In this study, we extend their analysis by incorporating $rs_{\rm cont}$ as an additional variable, along with the results for observed galaxies, as shown in Figures \ref{fig:peakshift} and \ref{fig:FWHM};
while $rs_{\rm cont}$ was fixed in the previous study, we explore a range of values here.

The main trends reported by \citet{Song2020} are well reproduced and not altered by variations in $rs_{\rm cont}$: the positive correlations of peak shift and FWHM with optical depth remain the strongest in our results, and the opposite dependence of peak shift and FWHM on the velocity structure is also confirmed.
To briefly repeat the discussion in the previous study, the positive correlations with optical depth are expected given the resonant nature of Ly$\alpha$ scattering, where a higher optical depth results in more scattering and thus leads to a larger peak shift and broader FWHM.
A larger medium velocity at a given radius (i.e., with a smaller $r_{\rm peak}$ and larger $V_{\rm peak}$) can Doppler-shift photons farther from the central wavelength in the medium frame, thereby suppressing scattering.
This explains the positive and negative correlations of peak shift with $r_{\rm peak}$ and $V_{\rm peak}$, respectively.
While FWHM shows similar correlations to peak shift, it also exhibits the reverse trends: negative correlation with $r_{\rm peak}$ and positive correlation with $V_{\rm peak}$.
This is because a smaller $r_{\rm peak}$ and a larger $V_{\rm peak}$ generate a broader velocity range within a given spatial range, which in turn broadens the wavelength distribution of photons in the medium frame, resulting in a larger FWHM as long as scattering remains efficient.

By considering a range of $rs_{\rm cont}$ values, we are able to examine its impact on the spectral features based on the distribution of the full model set.
$rs_{\rm cont}$ has a negligible impact on the peak shift, but a modest influence on the FWHM; specifically, a larger $rs_{\rm cont}$ leads to a larger FWHM, as photons are scattered over a wider range of velocities.
This is the same mechanism that leads to the $V_{\rm peak}$--FWHM correlation.
While the spatial extent of Ly$\alpha$ halos is shaped by a complex interplay among the extents of the medium and the source as well as the medium velocity structure, peak shift and FWHM are predominantly governed by optical depth.
Nevertheless, the substantial scatter in the correlations with optical depth suggests that these features are modulated by other parameters, once again underscoring the multi-dimensional nature of Ly$\alpha$ radiative transfer.

\begin{figure}[t]
   \centering
   \includegraphics[width=9cm]{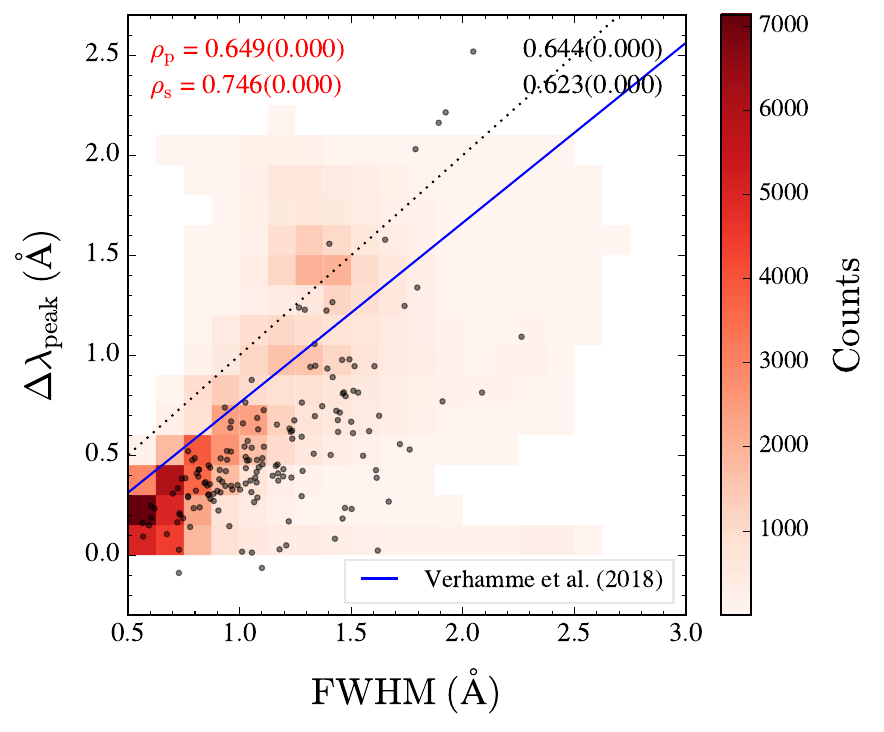}
   \caption{Correlation between peak shift and FWHM. The red 2D histogram is for the full set of models, and the gray dots represent the best-fit models for observed galaxies. The blue solid line corresponds to the empirical relation established by \citet{Verhamme2018}, and the dotted line corresponds to the $y=x$ line. The Pearson correlation coefficient ($\rho_{\rm p}$) and Spearman rank correlation coefficient ($\rho_{\rm s}$) are shown in red for all models and in black for the best-fit models, with the corresponding $p$-value in parentheses.}
   \label{fig:shift_fwhm}
\end{figure}

Lastly, we investigated the correlation between peak shift and FWHM, which has been proposed as a means to estimate systemic redshifts in the absence of lines other than Ly$\alpha$ \citep{Verhamme2018}.
Compared to Figure 10 in \citet{Song2020}, we included the models of varying $rs_{\rm cont}$ as well as the subset of best-fit models in Figure \ref{fig:shift_fwhm}.
Consistent with \citet{Song2020} and the strong positive correlations of peak shift and FWHM with optical depth as seen in Figures \ref{fig:peakshift} and \ref{fig:FWHM}, we find a tight correlation between peak shift and FWHM in both the full model set and the best-fit subset.
While our results are systematically offset from the empirical relation reported in \citet{Verhamme2018}, this offset is not necessarily problematic, as the sample used in \citet{Verhamme2018}--consisting of $\sim45$ galaxies--may also be subject to sample bias.
Specifically, their sample was designed to include galaxies with double-peaked Ly$\alpha$ profiles, whereas our sample consists solely of galaxies with red peak-only profiles, likely indicating a larger velocity in the expanding medium.
Recalling the dependence of peak shift and FWHM on the medium velocity shown in Figures \ref{fig:peakshift} and \ref{fig:FWHM}--namely, that higher velocities result in smaller peak shifts and larger FWHMs--galaxies with higher (lower) medium velocities tend to be distributed toward the lower right (upper left) in the peak shift--FWHM plane.
This may explain why our sample and that of \citet{Verhamme2018} occupy different regions in the plane (located in the lower right and in the upper left, respectively).
Given the large variation across samples and models, this correlation may not be sufficiently robust to be used for estimating systemic redshifts.
However, with prior knowledge of the medium velocity profile, it may be possible to tailor the correlation to be tighter and applicable to specific cases.

\section{Summary}\label{sec:sum}
We conducted Ly$\alpha$ radiative transfer modeling for Ly$\alpha$ halos around 163 star-forming galaxies at $z=$3--6 observed by MUSE, to reproduce their spectra and SBPs.
To this end, we utilized the Ly$\alpha$ radiative transfer calculations by \citet{Song2020}, which assume an outflowing halo with an extended source distribution.
The spectra and SBPs of Ly$\alpha$ halos of our sample galaxies were generally well reproduced using a model with a broad density distribution and low expanding velocity at large radii.
This success in reproducing a larger galaxy sample further supports the validity of this simple model.
Additionally, the necessity of modeling both the spectrum and SBP simultaneously was confirmed with this expanded dataset.
Our correlation analysis between observables and model parameters reveals that Ly$\alpha$ emission is governed by multiple parameters in a complex and interdependent way.
While the extents of the medium and the source and the optical depth emerge as the primary factors shaping the extent of Ly$\alpha$ halos and spectral shapes, respectively, the medium velocity structure can significantly alter these primary correlations, leading to substantial scatter across different galaxy samples.

\begin{acknowledgments}
The authors are grateful to the referee for his/her insightful comments and to Floriane Leclercq for generously sharing the MUSE IFU data and for her kind assistance. 
This work was supported by the National Research Foundation of Korea (NRF) grant funded by the Korean government (MSIT) (No. 2022M3K3A1093827, 2022R1A4A3031306).
K.I.S was supported by a NRF grant funded by the MSIT (No. 2020R1A2C1005788) and by the Korea Astronomy and Space Science Institute grant funded by the MSIT (No. 2025186902).
\end{acknowledgments}

\bibliography{ref}{}
\bibliographystyle{aasjournal}
\end{document}